\documentclass[aps,preprint]{revtex4}%
\usepackage{amsfonts}
\usepackage{amsmath}
\usepackage{amssymb}
\usepackage{graphicx}%
\begin{document}
\title{Multiple Quantum NMR Dynamics in Dipolar Ordered Spin Systems}
\author{S. I. Doronin, E. B Fel'dman, E. I. Kuznetsova}
\affiliation{Institute of Problems of Chemical Physics of Russian Academy of Sciences,
Chernogolovka, Moscow Region, 142432, Russia}
\author{G. B. Furman, S.D. Goren}
\affiliation{Department of Physics, Ben Gurion University, Beer Sheva 84105, Israel}

\begin{abstract}
We investigate analytically and numerically the Multiple Quantum (MQ) NMR
dynamics in systems of nuclear spins 1/2 coupled by the dipole-dipole
interactions in the case of the dipolar ordered initial state. We suggest two
different methods of MQ NMR. One of them is based on the measurement of the
dipolar temperature in the quasi-equilibrium state which establishes after the
time of order T2 after the MQ NMR experiment. The other method uses an
additional resonance 45\symbol{94}0-pulse after the preparation period of the
standard MQ NMR experiment in solids. Many-spin clusters and correlations are
created faster in such experiments than in the usual MQ NMR experiments and
can be used for the investigation of many-spin dynamics of nuclear spins in solids

\end{abstract}
\startpage{1 }
\maketitle

\section{Introduction}

Multipe-quantum (MQ) NMR spin dynamics in solids \cite{baun1985} is a powerful
tool for the investigation of structure and dynamical processes in solids,
counting the number of spins in impurity clusters \cite{baum1986} and the
simplification of ordinary NMR spectra \cite{warren1980}. Although MQ NMR was
successful in a lot of applications and experimental methods of MQ NMR have
been developed adequately, the theoretical interpretation of many-spin MQ NMR
dynamics is restricted by the phenomenological approach \cite{baun1985}. A
systematic quantum-mechanical approach was developed
\cite{feldman1997,doronin2000,doronin2005} only for one-dimensional systems in
the approximation of nearest neighbor dipolar interactions. Up to now the
thermodynamic equilibrium density matrix in a strong external magnetic field
has been considered as the initial condition for these experiments and
theoretical interpretations. Recently, it has been suggested \cite{furman2005}
to consider the dipolar ordered state as the initial state for such
experiments. It is well known that the dipolar ordered state can be prepared
using the method of adiabatic demagnetization in a rotating frame (ADRF)
\cite{goldman1970,slichter1961} or with the Jeener-Broekaert (JB) two-pulse
sequence \cite{goldman1970,jeener1967}. It has been shown that the spin system
contains MQ coherences immediately following the second pulse of the JB pulse
sequence \cite{Emid1979}. By encoding the coherence numbers in an orthogonal
basis (the $x$-basis), it was shown that the dipolar ordered state is a
two-spin correlated one \cite{Cory2003}. As a result of using the dipolar
ordered initial condition, many-spin clusters and correlations appear faster
than in the ordinary MQ NMR experiments in solids \cite{baun1985} and some
peculiarities of MQ dynamics can be investigated with these experiments. Of
course, it is necessary to make some changes in the scheme of the standard
experiment in order to obtain non-zero signals of MQ coherences.

In the present article we consider MQ NMR dynamics when the initial condition
is determined by the dipolar ordered state. Our motivation of performance of
this work is defined first of all by that the many-spin correlations are
created faster in such experiments and can be used for the investigation of
many-spin dynamics of nuclear spins in solids.We consider peculiarities of MQ
NMR in systems prepared in the dipolar ordered states in Section II. Two
methods of MQ NMR for the systems in the dipolar ordered states are presented
in Sections III and\ IV. The method of Section III is based on the measurement
of the dipolar temperature of the quasi-equilibrium state which establishes
after the time $t\sim\omega_{loc}^{-1}$ ($\omega_{loc}$ is the dipolar local
field) after the MQ NMR experiment. The method of Section IV is a simple
modification of the usual MQ NMR experiment \cite{baun1985}. The additional
$(\pi/4)_{y}$-pulse after the preparation period allows the observation of a
non-zero signal as in the usual MQ NMR experiments. Computer simulations of
such experiments for linear chains containing up to eight spins are presented
in Section V.

\section{MQ NMR with the initial dipolar ordered state}

We consider a system of nuclear spins ($s=1/2$) coupled by the dipole-dipole
interaction (DDI) in a strong external magnetic field. The secular part of the
DDI Hamiltonian \cite{goldman1970} has the following form
\begin{equation}
{\mathcal{H}}_{dz}=\sum_{j<k}D_{jk}\left[  I_{jz}I_{kz}-\frac{1}{4}\left(
I_{j}^{+}I_{k}^{-}+I_{j}^{-}I_{k}^{+}\right)  \right]  , \tag{1}%
\end{equation}
where $D_{jk}=\frac{\gamma^{2}\hslash}{r_{jk}^{3}}(1-3\cos^{2}\theta_{jk})$ is
the coupling constant between spins $j$ and $k$, $\gamma$ is the gyromagnetic
ratio, $r_{jk}$ is the distance between spins $j$ and $k$, $\theta_{jk}$ is
the angle between the internuclear vector $\overrightarrow{r_{jk}}$ and the
external magnetic field $\overrightarrow{H_{0}}$ which is directed along the
axis $z$. $I_{j\alpha}(\alpha=x,y,z)$ is the projection of the angular spin
momentum operator on the axis $\alpha$; $I_{j}^{+}$ and $I_{j}^{-}$ are the
raising and lowering operators of spin~$j$.

The basic scheme of the standard MQ NMR experiments consists of four distinct
periods of time (Fig.1a): preparation $\left(  \tau\right)  $, evolution
$\left(  t_{1}\right)  $, mixing $\left(  \tau\right)  $ and detection
$\left(  t_{2}\right)  $ \cite{baun1985}. MQ coherences are created by the
multipulse sequence consisting of eight-pulse cycles on the preparation period
~\cite{baun1985}. In the rotating reference frame (RRF) \cite{goldman1970} the
average Hamiltonian describing the MQ dynamics at the preparation period can
be written as \cite{baun1985}
\begin{equation}
\mathcal{H}_{MQ}=\mathcal{H}^{(2)}+\mathcal{H}^{(-2)}, \tag{2}%
\end{equation}
where $H^{(\pm2)}=-\frac{1}{4}\sum_{j<k}D_{jk}I_{j}^{\pm}I_{k}^{\pm}$. \ Then
the evolution period without any pulses follows . The transfer of the
information about MQ coherences to the observable magnetization occurs during
the mixing period. The density matrix of the spin system, $\rho(\tau)$, at the
end of the preparation period is
\begin{equation}
\rho(\tau)=U(\tau)\rho(0)U^{+}(\tau), \tag{3}%
\end{equation}
where $U(\tau)=\exp(-i\tau(H^{(2)}+H^{(-2)}))$ is the evolution operator on
the preparation period and $\rho(0)$ is the initial density matrix of the
system. On the mixing period the spin system is irradiated with sequence of
the pulses shifted on a $\pi/2-$phase regarding the pulses of the preparation
period. As a result the average Hamiltonian describing evolution of the spin
system on the mixing period changes a sign on opposite to the sign of the
Hamiltonian (2), and the evolution operator , $U(\tau),$ is replaced by the
operator $U^{+}(\tau).$\ Usually the thermodynamical equilibrium density
matrix is used as the initial one for MQ NMR experiments. Here we consider MQ
NMR dynamics with the initial dipolar ordered state when the Hamiltonian of
the system is determined by Eq. (1). Then the equilibrium state can be
described as
\begin{equation}
\rho(0)=\dfrac{1}{Z}\exp(-\beta\mathcal{H}_{dz})\thickapprox\left(
1-\beta\mathcal{H}_{dz}\right)  /2^{N}, \tag{4}%
\end{equation}
where $\beta$ is the inverse spin temperature, the partition function
$Z=Tr\{\exp(-\beta H_{dz})\}$ and $N$ is the number of spins in the system.
The high temperature approximation is taken into account in Eq. (4). The
dipolar ordered state of the spin system can be reached by using the method of
adiabatic demagnetization in a rotating frame (ADRF)
\cite{goldman1970,slichter1961} or by applying a pair of phase-shifted
radiofrequency (rf)-pulses (the Jeener-Brokaert (JB) method)
\cite{goldman1970,jeener1967}. It is evident that the unit operator in Eq. (4)
is not significant for the time evolution of the density matrix. For
simplicity we will take
\begin{equation}
\rho(0)=\mathcal{H}_{dz} \tag{5}%
\end{equation}
as the initial condition. The use of the dipolar ordered state as the initial
condition in MQ NMR experiments leads to the emergence of spin clusters with a
greater number of spins faster than in the standard method. It is well known
\cite{goldman1970} that NMR methods in solids use both the Zeeman equilibrium
state and the dipolar ordered one as the initial conditions. Here the
developed methods together with the standard one \cite{baun1985} yield
analogous possibilities for MQ NMR spectroscopy. Below we describe two
different methods of MQ NMR of the dipolar ordered systems.

\section{The Dipolar Temperature as a Source of Information about MQ NMR}

Starting with the initial condition (5), the density matrix, $\rho(t)$, after
the three periods of the standard MQ NMR experiment (Fig.1a) can be written as%

\begin{equation}
\rho(t)=U^{+}(\tau)e^{-i\delta t_{1}I_{z}}\rho(\tau)e^{i\delta t_{1}I_{z}%
}U(\tau), \tag{6}%
\end{equation}
where $\rho(\tau)$\ is the density matrix at the end of the preparation period
according to Eq. (3) and $t=2\tau+t_{1}$, $\delta$ is the frequency offset on
the evolution period of the duration $t_{1}$ which is a result of applying the
time proportional phase incrementation (TPPI) method \cite{baun1985}. The last
unitary transformation in (6) with operator $U(\tau)$\ describes the mixing
period. Then the evolution of the system in the RRF is governed by the
Hamiltonian $H_{dz}$ of Eq. (1) and after the time $t\geq\omega_{loc}^{-1}$
$\left(  \omega_{loc}^{2}=Tr\{\mathcal{H}_{dz}^{2}\}/Tr\{I_{z}^{2}\}\right)  $
the system achieves the two-temperature quasi-equilibrium state with the
density matrix $\rho_{eq}$,
\begin{equation}
\rho_{eq}=\alpha\omega_{0}I_{z}+\beta\mathcal{H}_{dz}, \tag{7}%
\end{equation}
where the inverse temperatures, $\alpha$ and $\beta$, can be found from the
conservation laws:
\begin{equation}
\alpha=\frac{Tr\{\rho(t)I_{z}\}}{\omega_{0}Tr\{I_{z}^{2}\}},\;\;\beta
=\frac{Tr\{\rho(t)\mathcal{H}_{dz}\}}{Tr\{\mathcal{H}_{dz}^{2}\}}. \tag{8}%
\end{equation}
Applying the unitary transformation $V=e^{-i\frac{\pi}{2}I_{z}}e^{-i\pi I_{x}%
}$ to the expression $\rho(t)I_{z}$ one can obtain from Eq. (8) that
$\alpha=0$. It means that the quasi-equilibrium state of Eq. (7) is the
dipolar ordered state. It is convenient to expand the density matrix,
$\rho(\tau)$, at the end of the preparation period of the MQ NMR experiment
(Fig. 1a) with the initial condition (5) as \cite{feldman1997}
\begin{equation}
\rho(\tau)=\sum_{n}\rho_{n}(\tau), \tag{9}%
\end{equation}
where the term $\rho_{n}(\tau)$ is responsible for the MQ coherence of the
$n$-th order. One can find
\begin{equation}
e^{-i\delta I_{z}t}\rho_{n}(\tau)e^{i\delta I_{z}t}=e^{-in\delta t}\rho
_{n}(\tau). \tag{10}%
\end{equation}
By using Eqs. (8)-(10) one can rewrite the temperature $\beta$ as
\begin{equation}
\beta=\sum_{n}e^{-in\delta t}J_{n}(\tau), \tag{11}%
\end{equation}
where the intensities of MQ coherences
\begin{equation}
J_{n}(\tau)=\frac{Tr\{\rho_{n}(\tau)\rho_{-n}(\tau)\}}{Tr\{\mathcal{H}%
_{dz}^{2}\}}. \tag{12}%
\end{equation}

The temperature $\beta$ can be measured using the magnetization component
which is in phase with the exciting rf-pulse \cite{goldman1970}. Different
frequency components of the temperature $\beta$ yield MQ coherences of
different orders.

\section{The Modification of the standard MQ NMR Experiment in Solids}

In the previous section we showed that at the initial condition of Eq. (5) the
standard scheme of MQ NMR experiments does not lead to any observable
magnetization. In order to overcome this problem we introduce a $\phi_{y}%
$-pulse turning spins around the axis $y$ by the angle $\phi$ after the
preparation period (Fig. 1b). Then starting with $\rho(0)=H_{dz}$, after the
mixing period (Fig. 1b) the longitudinal magnetization, $M_{z}\left(
t\right)  $ , is
\begin{equation}
M_{z}\left(  t\right)  =Tr\left\{  \rho(t)I_{z}\right\}  \tag{13}%
\end{equation}
where the density matrix after the three periods of the modified MQ NMR pulse
sequence (Fig. 1b) can be written as
\begin{equation}
\rho(t)=U^{+}(\tau)e^{-i\delta I_{z}t_{1}}\rho_{\phi}(\tau)e^{i\delta
I_{z}t_{1}}U(\tau) \tag{14}%
\end{equation}
and $\rho_{\phi}(\tau)$ is the density matrix just after the $\phi_{y}%
$-pulse\ (Fig. 1b)%

\begin{equation}
\rho_{\phi}(\tau)=e^{-i\phi I_{y}}U(\tau)\rho(0)U^{+}(\tau)e^{i\phi I_{y}}.
\tag{15}%
\end{equation}
It is worth to notice that here the TPPI method \cite{baun1985} is applied in
the mixing period in contrast to the method of Section III when it was applied
in the preparation period of the MQ NMR experiment. Using Eqs. (14) and (15)
and the initial condition (5) it is convenient to present the formula (13) for
the longitudinal magnetization, $M_{z}\left(  t\right)  ,$ \ as follows%
\begin{equation}
M_{z}\left(  t\right)  =Tr\left\{  e^{-i\delta I_{z}t}e^{-i\phi I_{y}}%
U(\tau)H_{dz}U^{+}(\tau)e^{i\phi I_{y}}e^{i\delta I_{z}t}\rho_{MQ}%
(\tau)\right\}  \tag{16}%
\end{equation}
where
\begin{equation}
\rho_{_{MQ}}(\tau)=U(\tau)I_{z}U^{+}(\tau), \tag{17}%
\end{equation}
coincides with the density matrix at the end of the preparation period of the
\ standard MQ NMR experiment with the thermodynamical equilibrium density
matrix as the initial condition \cite{baun1985}. The density matrix
$\rho_{_{MQ}}(\tau)$ can be represented in the following form
\cite{feldman1997}
\begin{equation}
\rho_{_{MQ}}(\tau)=\sum_{n}\rho_{n}^{MQ}(\tau) \tag{18}%
\end{equation}
where again the term $\rho_{n}^{MQ}(\tau)$ is responsible for the MQ coherence
of the $n$-th order and satisfies to the relationship of Eq. (10). By using
Eqs. (10), (17), and (18) one can rewrite the expression for the observable
signal in terms of the intensities of MQ coherences
\begin{equation}
M_{z}\left(  t\right)  =-i\sum_{n}e^{-in\delta t}J_{n}(\tau)=i\sum
_{n}e^{in\delta t}J_{n}^{\ast}(\tau)=i\sum_{n}e^{-in\delta t}J_{-n}^{\ast
}(\tau), \tag{19}%
\end{equation}
where
\begin{equation}
J_{n}(\tau)=iTr\left\{  \rho_{\phi}(\tau)\rho_{n}^{MQ}(\tau)\right\}  \tag{20}%
\end{equation}
and the longitudinal magnetization, $M_{z}\left(  t\right)  $, is always real.
Thus, one can find from Eq.(19) that
\begin{equation}
J_{n}(\tau)=-J_{-n}^{\ast}(\tau). \tag{21}%
\end{equation}
One can notice that
\begin{equation}
I_{z}(\delta)=-I_{z}(-\delta) \tag{22}%
\end{equation}
by applying a $\pi$-rotation around the $y$-axis to all operators in Eq. (13).
It follows immediately from Eqs. (19) and (22) that
\begin{equation}
J_{n}(\tau)+J_{-n}(\tau)=0. \tag{23}%
\end{equation}
It is evident that with the method of Section III $J_{n}(\tau)=J_{-n}(\tau)$,
and the longitudinal magnetization, $M_{z}\left(  t\right)  $, is determined
as
\begin{equation}
M_{z}\left(  t\right)  =\sum_{n}\cos(n\delta t)J_{n}(\tau). \tag{24}%
\end{equation}
By using the modified pulse sequence (Fig. 1b) the longitudinal magnetization
can be expressed as
\begin{equation}
M_{z}\left(  t\right)  =-\sum_{n}\sin(n\delta t)J_{n}(\tau). \tag{25}%
\end{equation}
Thus the phases of the MQ NMR coherences in the modified MQ NMR pulse
excitation scheme are shifted by $\pi/2$ from the phases in standard MQ NMR
\cite{baun1985}. It is clear from Eq. (23) that the intensity of the MQ
coherence of the zeroth order, $J_{0}(\tau)$, is equal to zero in the
considered experiments.

It is well-known that in the usual MQ NMR experiments the sum of the
intensities of all MQ coherences does not depend on time \cite{lattrop1994}.
This is also right for the method described in Section III. Here this law has
a specific form. According to Eq. (23) the sum of intensities of orders $n$
and $-n$ is equal to zero for all $n$. Thus, the sum of the intensities of all
MQ coherences equals zero.

\section{The numerical analysis of the time evolution of MQ coherences}

We restrict ourself to numerical simulations of MQ NMR dynamics of
one-dimensional systems. For example, quasi-one-dimensional hydroxyl proton
chains in calcium hydroxyapatite $Ca_{5}(OH)(PO_{4})_{3}$ \cite{Cho1996} and
fluorine chains in calcium fluorapatite$Ca_{5}F(PO_{4})_{3}$ \cite{Cho1996}
are suitable objects to study MQ dynamics in dipolar ordered systems.The
numerical calculations were performed with the methods of Sections III, IV for
MQ NMR dynamics of linear chains of 6 and 8 spins. The DDI coupling constant
of the nearest neighbors is chosen to be $D=1s^{-1}$. Then the coupling
constant of spins $j$ and $k$ is $D/|j-k|^{3}$. In order to compare the
results of the numerical simulations with the analogous ones for the ordinary
MQ NMR dynamics we introduce the normalized intensities of MQ coherences for
the method of Section IV, $J_{n}(\tau)/\{Tr(I_{z}^{2})Tr(H_{dz}^{2})\}^{1/2}$.
The dependence of the intensities of MQ coherences on the dimensionless time,
$t=D\tau$ , of the preparation period in spin chains containing six and eight
spins is presented in Figs. 2 and 3 for the both methods of Sections III and
Sections IV. One can compare the intensities of MQ coherences in the suggested
experiments with the standard ones (see Figs. 2 and 3). It is evident that the
suggested methods can be considered as a useful addition to the standard MQ
NMR methods. Comparison of the suggested methods of MQ NMR of the systems in
the dipolar ordered state is given in Fig. 2 for MQ coherences of the second
and fourth orders. One can see (Figs. 2a, 2b) that the method of Section III
is more preferable for experimental realizations in some cases.

Fig. 2a demonstrates that the method of Section III yields the intensities of
MQ coherences of the fourth order which several times higher than the
analogous coherence in the standard MQ NMR experiment. It is clear from the
inset of Fig. 3a that MQ coherence of the sixth order, obtained with the
method of Section IV, appears little earlier than in the usual MQ NMR in a
linear chain of six spins. At the same time, the intensity of this MQ
coherence obtained with the method of Section III is equal zero. In fact, the
both methods supplement each other at the study of MQ NMR dynamics. A tendency
of the faster growth of MQ coherences of high orders takes place for the
linear chain containing eight spins (Fig. 3b). This peculiarity is connected
with the initial dipolar ordered state, $\rho(0)=H_{dz}$. As a result,
many-spin clusters and connected with them MQ coherences appear faster than in
the standard MQ NMR with the initial condition, $\rho(0)=I_{z}$. The numerical
calculations confirm also the results obtained in the previous section. In
particular, the growth of MQ coherences (for the method of Section IV) occurs
in accordance with Eq. (23) and $J_{0}(t)\equiv0$.

One can see from Figs. 2 and 3 and Eq. (23) that the observable intensities
can be negative in contrast to the ordinary MQ NMR experiments at high
temperatures \cite{baun1985,feldman1997,doronin2000}. At the same time, it was
shown \cite{feldman2002} that the intensities of MQ coherences can be negative
in the standard NMR experiments at low temperatures. It is a terminological
problem only. In fact, in MQ NMR experiments the observable is the
longitudinal magnetization modulated by rf pulses. The distinct frequency
components of the magnetization can have an arbitrary sign \cite{feldman2002}.

\section{Conclusions}

The MQ NMR methods for the detection of MQ coherences starting from the
dipolar ordered state are proposed. The first method is based on the
quasi-equilibrium state which establishes after the time $t\gtrsim\omega
_{loc}^{-1}$ after the MQ NMR experiment. This state is the dipolar ordered
one and its temperature is a source of information about MQ NMR dynamics. In
order to observe a non-zero signal of MQ coherences in the second method an
additional resonance $\left(  \pi/4\right)  _{y}$-pulse should be applied
after the preparation period of the standard MQ NMR experiment in solids.
Investigations of MQ NMR dynamics in the dipolar ordered states can be
considered as a supplementary method which complements the usual NMR in order
to study structures and dynamical processes in solids. Many-spin clusters and
many-spin correlations are created faster in such experiments than in the
usual MQ NMR with the initial condition without any correlation between the
spins. This paper focuses on simple one-dimensional examples but the physical
picture obtained here is not limited to the performed simulations and
experiments and opens new possibilities for the study of many-spin systems.

\subsection*{Acknowledgements}

We are grateful for financial support for this work through a grant from the
Russian Foundation for Basic Research (grant no. 07-07-00048). G.F.
acknowledges financial support from the US-Israel Binational Science
Foundation (grant no. 2002054).

\bigskip

Caption figures

Fig. 1 The basic schemes for MQ NMR: a) the scheme of the standard MQ NMR
experiment; b) the scheme of the modified MQ NMR experiment.

Fig. 2 The time dependence of the intensities of the MQ coherences of the
second order in a linear chain of six spins coupled by the DDI for
$\rho(0)=H_{dz}$; the intensities $J_{-2}$ (solid) and $J_{+2}$ (dash)
obtained by the method of Section IV at $\phi=\pi/4$. The intensity $J_{2}$
(dot) is obtained by the method of Section III. (b) The time dependence of the
intensities of the fourth order MQ coherences in a linear chain of six spins
coupled by the DDI : the intensity of the MQ coherence of the fourth order
(solid) for $\rho(0)=I_{z}$; $J_{-4}$ (dash) and $J_{+4}$ (dot) for
$\rho(0)=H_{dz}$ obtained by the method of Section IV at $\phi=\pi/4$, and the
intensity $J_{4}$ (dash-dot) is obtained by the method of Section III.

Fig. 3 The time dependence of the intensities of the sixth order MQ coherences
in a linear chain of six spins coupled by the DDI: the intensity of MQ
coherence of the sixth order (solid) for $\rho\left(  0\right)  =I_{z}$;
$J_{-6}$ (dash) for $\rho\left(  0\right)  =H_{dz}$ obtained by the method of
Section IV with using $(\pi/4)_{y}$-pulse and $J_{6}$(dot) obtained by the
method of Section III. (b) The time dependence of the intensities of the sixth
order MQ coherences in a linear chain of eight spins coupled by the DDI : the
intensity of MQ coherence of the sixth order (solid) for $\rho\left(
0\right)  =I_{z}$; $J_{-6}$ (dash) for $\rho\left(  0\right)  =H_{dz}$
obtained by the method of Section IV with $(\pi/4)_{y}$-pulse; $J_{6}$
(dot-dash) and $J_{8}$ (dot) obtained by the method of Section III. The insets
show that the MQ coherences in the dipolar ordered state obtained by the
method of Section IV with applied $(\pi/4)_{y}$-pulse appears little earlier
than in the usual MQ NMR and in the method of Section III.


\begin{thebibliography}{99}                                                                                               %


\bibitem {baun1985}J. Baum, M. Munovitz, A. N. Garroway, A. Pines, J. Chem.
Phys. \textbf{83} , 2015 (1985).

\bibitem {baum1986}J. Baum, K. K. Gleason, A. Pines, J. Chem. Phys.,
\textbf{56}, 1377 (1986).

\bibitem {warren1980}W. S. Warren, D. P. Weitekamp, A. Pines, J. Chem. Phys.,
\textbf{73}, 2084 (1980).

\bibitem {feldman1997}E. B. Fel'dman, S. Lacelle, J. Chem. Phys.,
\textbf{107}, 7067 (1997).

\bibitem {doronin2000}S. I. Doronin, I. I. Maximov, E. B. Fel'dman, Zh. Eksp.
Teor. Fiz., \textbf{118}, 687 (2000), JETP, \textbf{91}, 597 (2000).

\bibitem {doronin2005}S. I. Doronin, E. B. Fel'dman, Solid State Nucl. Magn.
Reson., \textbf{28}, 111 (2005).

\bibitem {furman2005}G. B. Furman, S. D. Goren, J. Phys.: Condens. Matter,
\textbf{17}, 4501 (2005).

\bibitem {goldman1970}M. Goldman. \textit{Spin Temperature and Nuclear
Magnetic Resonance in Solids}, Oxford. Clarendon Press. 1970.

\bibitem {slichter1961}C. P. Slichter, W. C. Holton, Phys. Rev., \textbf{122},
1701 (1961).

\bibitem {jeener1967}J. Jeener, P. Broekaert, Phys. Rev., \textbf{157}, 232 (1967).

\bibitem {Emid1979}S. Emid, A. Bax, J. Konijnendijk, J. Smidt, A. Pines,
Physica B, \textbf{96}, 333 (1979).

\bibitem {Cory2003}H. Cho, D. G. Cory, C. Ramanathan, J. Chem. Phys.,
\textbf{118}, 3686 (2003).

\bibitem {lattrop1994}D. A. Lathrop, E. S. Handy, K. K. Gleason, J. Magn.
Reson. A , \textbf{111}, 161 (1994).

\bibitem {Cho1996}G. Cho, J. P. Yesinowski, J. Phys. Chem., \textbf{100},
15716 (1996).

\bibitem {feldman2002}E. B. Fel'dman, I. I. Maximov, J. Magn. Reson.,
\textbf{157}, 106 (2002).
\end{thebibliography}
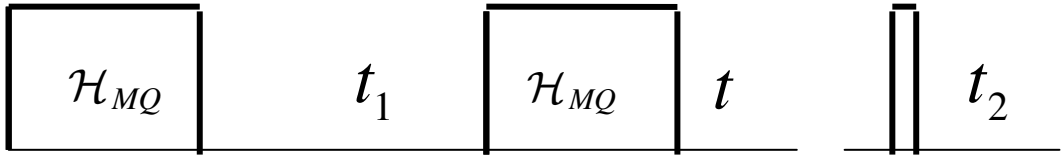
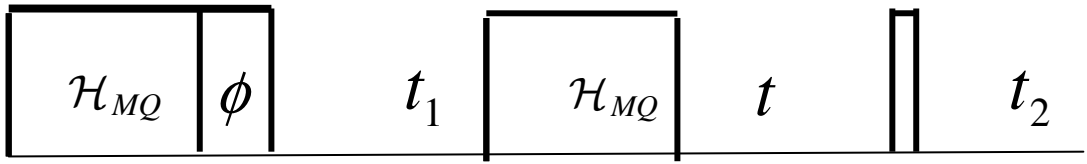
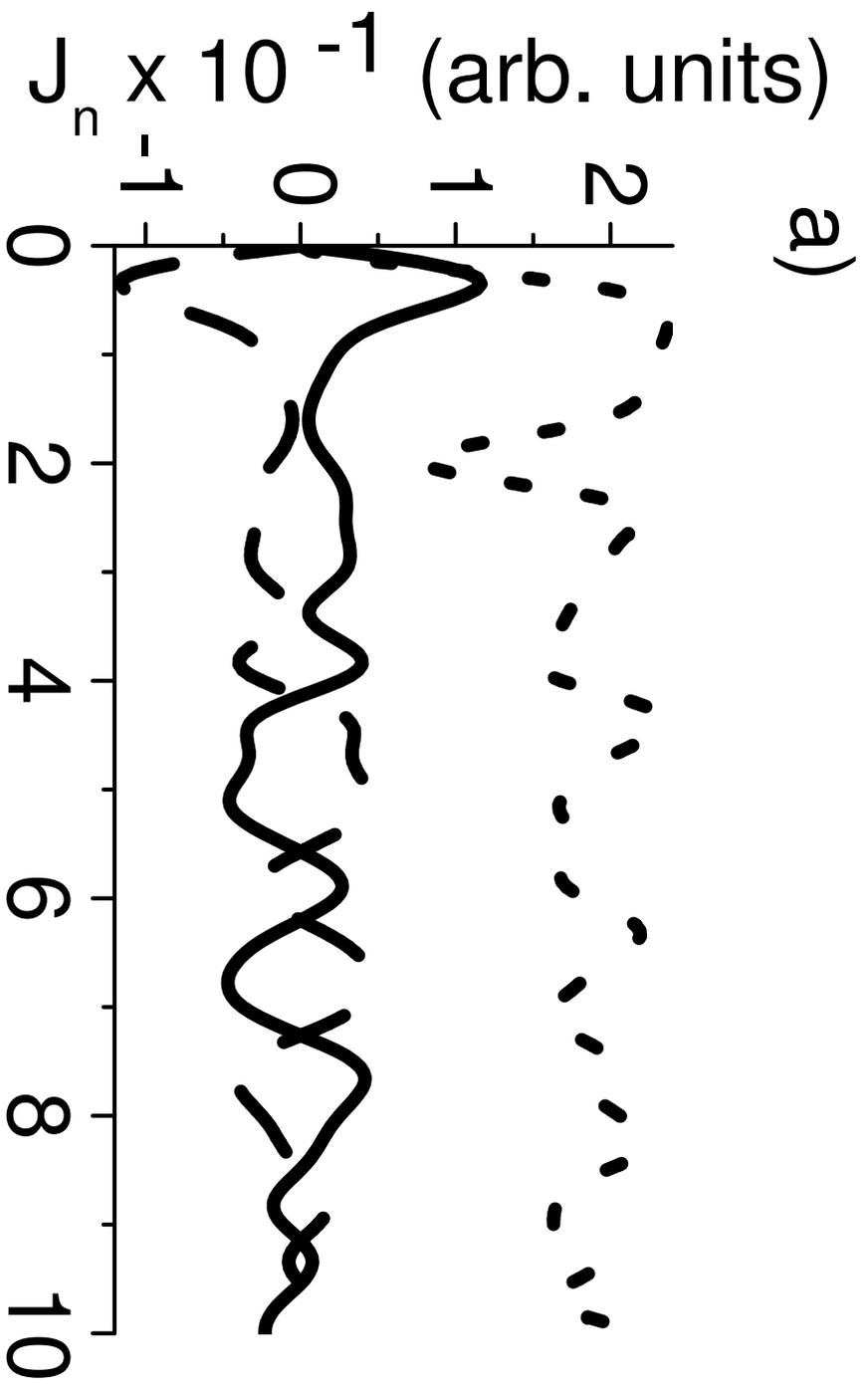
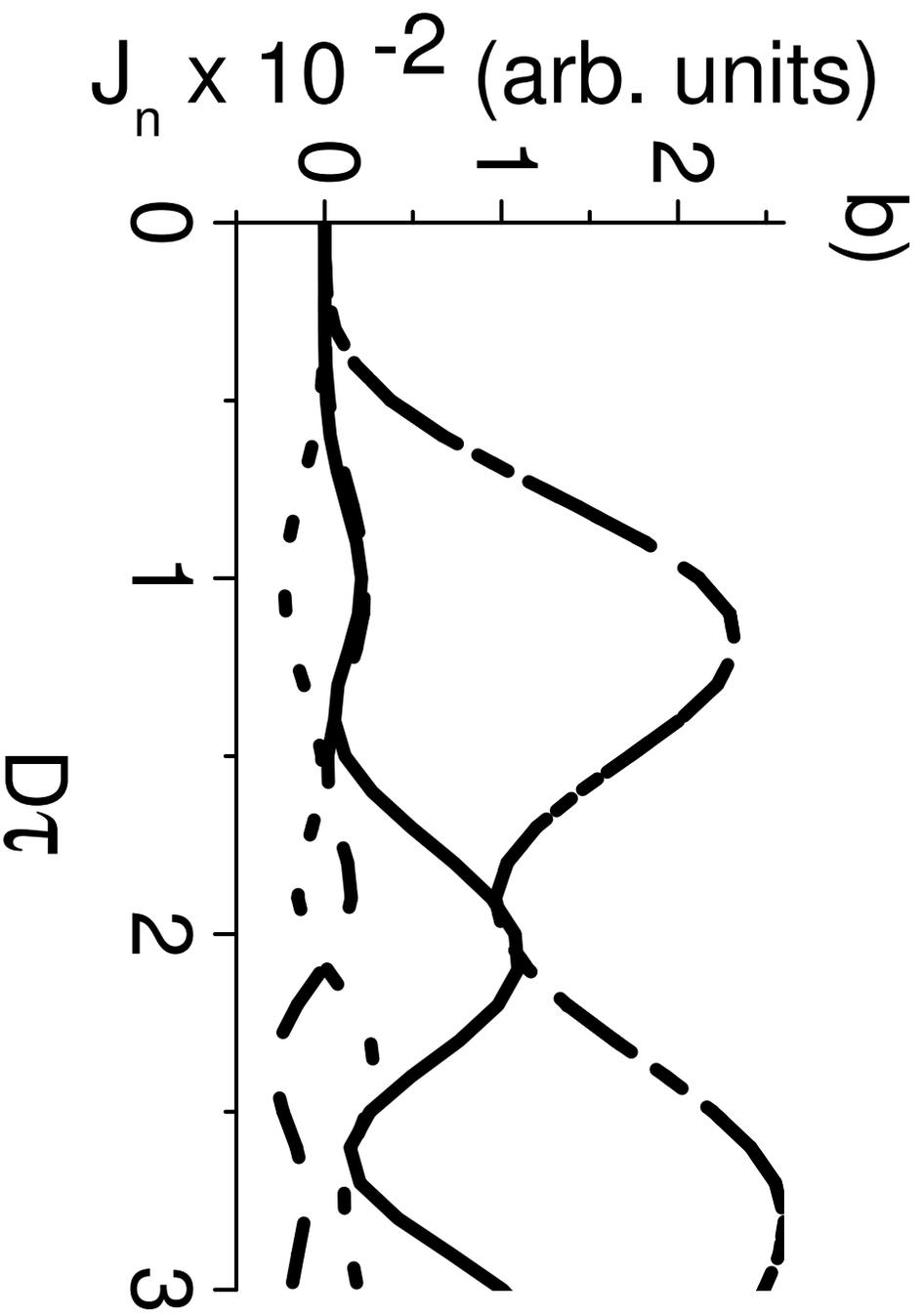
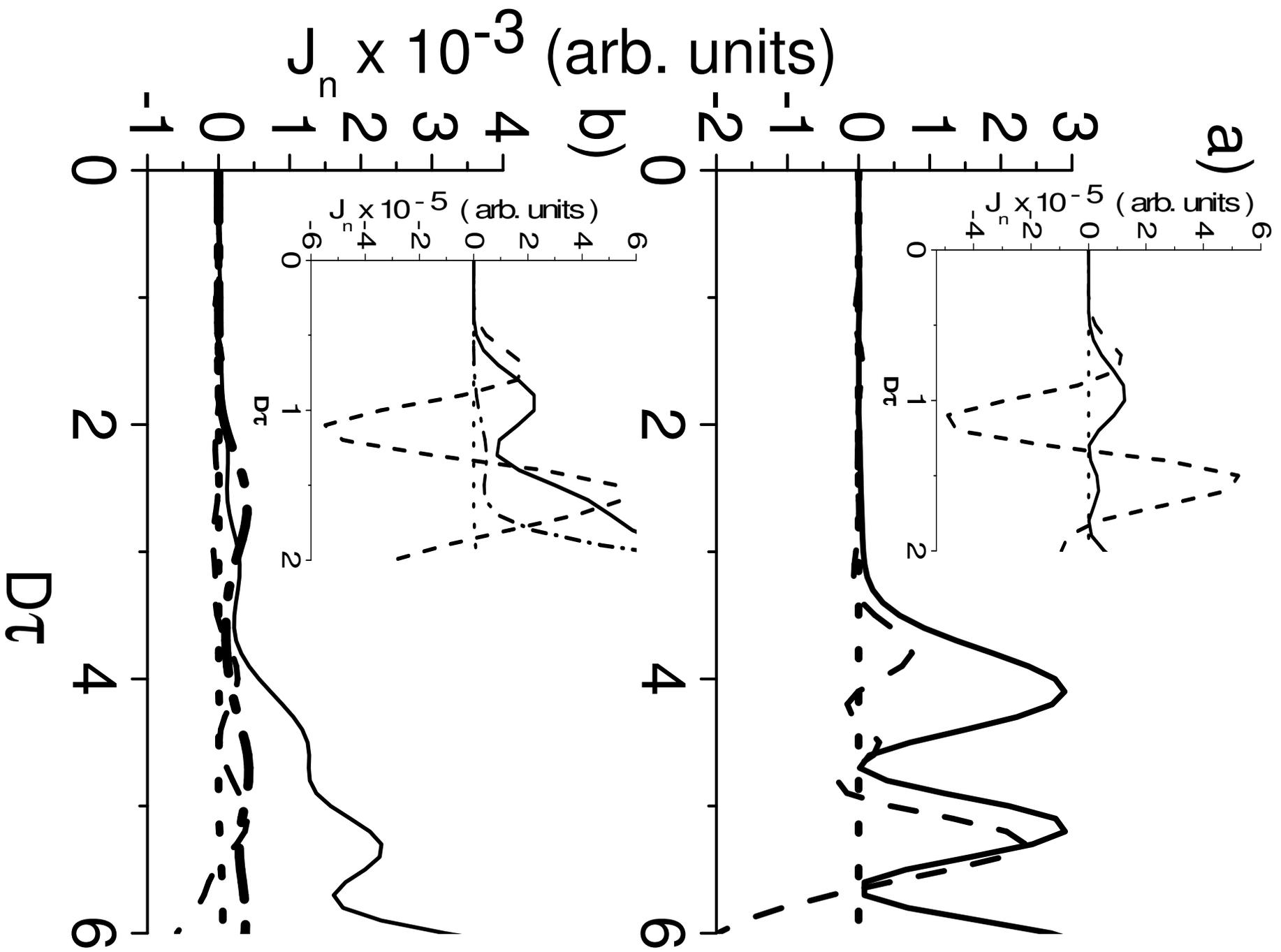
\end{document}